\documentclass[reprint,showkeys,
superscriptaddress,
 amsmath,
 amssymb,
 aps,
]{revtex4-2}

\usepackage[utf8]{inputenc}

\usepackage{graphicx}
\usepackage{dcolumn}
\usepackage{bm}
\usepackage[colorlinks]{hyperref}
\hypersetup{
    linkcolor = blue,   
    citecolor = blue,   
    urlcolor  = blue    
}
\usepackage{color}
\usepackage{amsmath}

\begin{document}

\title{Dynamical control of particle jets from a driven condensate in a one-dimensional lattice with double-well potential}

\author{Z. Li}
\affiliation{Basic Teaching Department, Nanhang Jincheng College, Nanjing 211156, China}

\author{L. Q. Lai}
\email{lqlai@njupt.edu.cn}
\affiliation{School of Science, Nanjing University of Posts and Telecommunications, Nanjing 210023, China}

\date{\today}

\begin{abstract}

We investigate the nonlinear dynamics of a Bose-Einstein condensate trapped in a double-well potential of a one-dimensional lattice, where the interatomic interactions are periodically modulated in time. In the typical case of a symmetric double-well, we observe collective particle emission under resonant driving, where the excitation regimes are explicitly constrained by the interplay between the drive strength and the hopping amplitude. By introducing a depth asymmetry between the wells, we find that moderate bias specifically enhances the emission rate, while large asymmetry suppresses it. The particle jets can be further controlled by modulating the hopping amplitudes, where the emission is weakened for finite hopping imbalances. These results outline the roles of asymmetry and external driving in precisely manipulating quantum many-body transport, and may offer insights into the design of atomtronic devices.

\end{abstract}


\maketitle

\section{Introduction}

The quantum dynamics of ultracold atomic gases has been attracting intensive attention for the past few decades since the realization of Bose-Einstein condensation \cite{bloch}, as it constitutes a versatile platform for concrete investigations of many-body phenomena relevant to condensed matter physics. In particular, the ability of precisely controlling the interatomic interactions via Feshbach resonance enables explorations of exotic quantum matter and quantum engineering \cite{chin}. Among the prominent examples, the system of bosonic atoms confined in a symmetric double-well potential is straightforwardly related to the celebrated Josephson effect in superconductors, popularly known as the bosonic Josephson junction \cite{gati}, which has been extensively investigated both theoretically and experimentally \cite{walls,ananikian,smerzi0,shin1,giovanazzi,diaz,qian,susanto,zhang,gu,hou,ying,sicks,hamza}.

Such a system serves as a paradigm model for studying various fundamental quantum features and offering insights into rich nonlinear dynamics that are not accessible for conventional superconducting junctions, including Josephson oscillations \cite{pitaevskii,shin2,smerzi1,salgueiro,wang,adhikari,abad,nieuwkerk}, macroscopic quantum self-trapping \cite{gillet,zibold,roy,smerzi2,jezek} and tunneling dynamics \cite{albiez,zollner,zollner1,zollner2,maraj,lai0,lai1}. The advances of precision measurement techniques have further established these systems as powerful quantum simulators for exploring many-body dynamics in optical lattices under tunable conditions \cite{cataliotti,morsch,mistakidis}.

The scenario becomes particularly intriguing in an asymmetric double-well configuration, where the explicit breaking of spatial symmetry gives rise to qualitatively new physics \cite{schumm,theocharis,hall,jezek1,kim,gavrilov,cosme,haldar,lindberg,korshynska}. The asymmetry typically induces significant changes in the dynamical behavior, and is associated with symmetry-breaking quantum phase transitions \cite{trenkwalder}. From an experimental perspective, an asymmetric double well can be realized as a fundamental building block of tilted optical lattices, which has been vital to the recent studies of resonantly enhanced tunneling \cite{rubbo,sias,sias1,buyskikh,alon}. Moreover, even a slight asymmetry, often unavoidable in realistic trapping potentials, can  influence various properties of the system in nontrivial ways, such as dynamical instabilities, population imbalance and tunneling rates \cite{lindberg,korshynska}, facilitating verifications of the crucial role of the interplay between interactions and asymmetries in quantum transports.

In the previous studies, we uncovered a variety of particle-emission phenomena in a parametrically driven Bose-Einstein condensate, including resonantly enhanced emission \cite{lai0}, drive-imbalance effects \cite{lai1}, interference-induced suppression of the emission \cite{lai2} and intermittent jet formation \cite{lai3}, while mainly focusing on the symmetric lattice geometries with homogeneous hopping amplitudes. In this work, we theoretically investigate the nonlinear dynamics within a one-dimensional lattice, where an asymmetric double-well potential is specifically applied to trap the condensate and the interatomic interactions are periodically modulated. We identify the distinct excitation regimes, and further demonstrate tunable control of collective particle jets via well asymmetries and hopping imbalances.

The paper is organized as follows. In Sec.~\ref{model}, we introduce the lattice model and illustrate a brief outline of the framework. In Sec.~\ref{parametric}, we present the numerical analysis of nonlinear dynamics and compare the results with analytical solutions. We summarize our work and give some concluding remarks in Sec.~\ref{conclusion}.

\section{Theoretical model}\label{model}

\begin{figure}[t]
\includegraphics[width=\columnwidth]{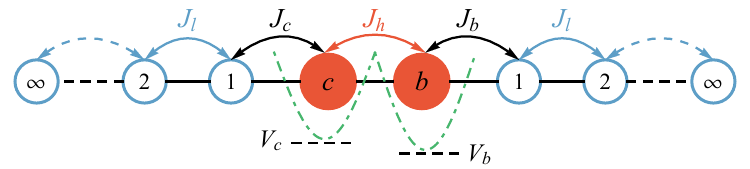}
\caption{Sketch of the one-dimensional infinite lattice under consideration. A double-well potential of depth $V_{b}$ and $V_{c}$ is applied to the central lattice sites labeled $b$ and $c$, and the blue empty circles indexed by integers $1,2,\ldots,\infty$ denote the remaining sites outside the wells.}
\label{dwlattice}
\end{figure}

As schematically shown in Fig.~\ref{dwlattice}, we consider a one-dimensional infinite lattice featuring a double-well potential that traps a Bose-Einstein condensate in the central lattice sites labeled $b$ and $c$, and the remaining sites in the leads are symbolized by nonzero integers. The coupling between the central sites is quantified by strength $J_{h}$, which facilitates the back-and-forth tunneling of atoms. Excited particles with sufficient energy can escape from the wells and hop to the nearest-neighboring sites with hopping amplitude $J_{b}$ and $J_{c}$, respectively, while traveling along the leads to infinity with a tunneling strength $J_{l}$. Since the atomic density and the resulting particle jets are low outside the wells when compared with the condensate, it is reasonable to only include the interatomic interactions in the central sites. The system can thus be mathematically described by the Hamiltonian
\begin{eqnarray}
\hat{H} &=& V_{b}\hat{b}_{0}^{\dagger}\hat{b}_{0}+V_{c}\hat{c}_{0}^{\dagger}\hat{c}_{0}-J_{h}(\hat{b}_{0}^{\dagger}\hat{c}_{0}+\hat{c}_{0}^{\dagger}\hat{b}_{0}) \nonumber \\
&&+\frac{G\left(t\right)}{2}\left(\hat{b}_{0}^{\dagger}\hat{b}_{0}^{\dagger}\hat{b}_{0}\hat{b}_{0}+\hat{c}_{0}^{\dagger}\hat{c}_{0}^{\dagger}\hat{c}_{0}\hat{c}_{0}\right) \nonumber \\
&&-J_{b}\left(\hat{b}_{0}^{\dagger}\hat{b}_{1}+\hat{b}_{1}^{\dagger}\hat{b}_{0}\right)
-J_{c}\left(\hat{c}_{0}^{\dagger}\hat{c}_{1}+\hat{c}_{1}^{\dagger}\hat{c}_{0}\right) \nonumber \\
&&-J_{l}\sum_{j=1}^{\infty}\left(\hat{b}_{j+1}^{\dagger}\hat{b}_{j}+\hat{c}_{j+1}^{\dagger}\hat{c}_{j}+{\rm{H.c.}}\right).
\end{eqnarray}
Here, $V_{b}$ and $V_{c}$ represent the respective depth of the double wells. $\hat{b}_{0}^{\dagger}$ $(\hat{b}_{0})$ and $\hat{c}_{0}^{\dagger}$ $(\hat{c}_{0})$ are the bosonic creation (annihilation) operators of the central sites, while $\hat{b}_{j}^{\dagger}$ $(\hat{b}_{j})$ and $\hat{c}_{j}^{\dagger}$ $(\hat{c}_{j})$ correspond to the $j$th site on the right or left leads. The time-dependent term $G(t)=U+g\left(t\right)$ characterizes the on-site pairwise interactions, where $U$ is a constant and $g(t)=g\sin(\omega t)$ is a sinusoidally oscillating driving, with $g$ being the drive strength and $\omega$ being the drive frequency.

We study the dynamics under the mean-field approximation, such that the operators can be approximately replaced by their expectation values with complex numbers $\mu_{j}=\langle \hat{\mu}_{j}\rangle\equiv\varphi_{\mu,j}$ ($\mu=\{b,c\}$), and physically $|\varphi_{\mu,j}|^{2}$ represents the particle number on the $j$th site. The time evolution of the system is governed by the Hamiltonian, which yields the corresponding Heisenberg equations of motion for site $j=0$ ($\hbar=1$ throughout)
\begin{eqnarray}
i\partial_{t}\varphi_{b,0} &=& V_{b}\varphi_{b,0}+G\left(t\right)|\varphi_{b,0}|^{2}\varphi_{b,0}-J_{h}\varphi_{c,0}-J_{b}\varphi_{b,1}, \label{b0} \\
i\partial_{t}\varphi_{c,0} &=& V_{c}\varphi_{c,0}+G\left(t\right)|\varphi_{c,0}|^{2}\varphi_{c,0}-J_{h}\varphi_{b,0}-J_{c}\varphi_{c,1}, \label{c0}
\end{eqnarray}
and for the remaining sites $j\geq1$,
\begin{eqnarray}
i\partial_{t}\varphi_{\mu,1} &=& -J_{\mu}\varphi_{\mu,0}-J_{l}\varphi_{\mu,2}, \\
i\partial_{t}\varphi_{\mu,j} &=& -J_{l}(\varphi_{\mu,j-1}+\varphi_{\mu,j+1}). \label{linear}
\end{eqnarray}
In cold atom experiments, the dc component of the scattering length is generally kept small, and a finite interaction $U$ does not qualitatively affect the relevant physics \cite{lai2,lai3,clark,clark1}; thus, we work in the limit of $U=0$ to further simplify the analysis (see Appendix \ref{finiteU}). In the absence of an external periodic driving ($g=0$), the system remains in equilibrium, and can be explicitly described by the two-mode approximation \cite{salgueiro,ananikian}.

We begin with the extensively studied case of equivalent depth of the wells ($V_{b}=V_{c}$) and introduce the stationary ansatz $
b_{0}=\beta e^{-i\epsilon_{b}t}$ and $c_{0}=\gamma e^{-i\epsilon_{c}t}$ ($\beta$ and $\gamma$ are constant), which naturally recovers the typical scenario with $\epsilon_{b}=\epsilon_{c}$,  reaching
\begin{eqnarray}
\left(
\begin{array}{cc}
 \Lambda_{b}   &  J_{h} \\
   J_{h}  &  \Lambda_{c}
\end{array}
\right)
\left(
\begin{array}{cc}
     \beta  \\
     \gamma
\end{array}
\right)=0
\end{eqnarray}
with $\Lambda_{\mu}=\epsilon_{\mu}-V_{\mu}-J_{\mu}^{2}\mathcal{G}_{11}(\epsilon_{\mu})$, and as derived in Appendix~\ref{Green's function},
\begin{eqnarray}
\mathcal{G}_{11}\left(\epsilon\right)=\frac{\epsilon}{2J_{l}^{2}}-i\sqrt{\frac{1}{J_{l}^{2}}-\frac{\epsilon^{2}}{4J_{l}^{4}}}
\end{eqnarray}
is the frequency-domain Green's function. When involving the asymmetric double well ($V_{b}\neq V_{c}$), a perturbative solution can be readily obtained with weak drive strength $g$ and weak hopping amplitude $J_{\mu}$. For the case of $\beta=\gamma$, to the zeroth order in $J_{\mu}$ we have
\begin{eqnarray}
\epsilon_{\mu,{\rm s}}^{(0)}=V_{\mu}-J_{h},
\end{eqnarray}
while for the antisymmetric case of $\beta=-\gamma$ it gives
\begin{eqnarray}
\epsilon_{\mu,{\rm as}}^{(0)}=V_{\mu}+J_{h}.
\end{eqnarray}
To the second order in the hopping amplitude $J_{\mu}$, a slick way involves the direct substitution of the zeroth-order solutions into the above matrix equation, leading to
\begin{eqnarray}
\epsilon_{\mu,{\rm s}}^{(2)}=\epsilon_{\mu,{\rm s}}^{(0)}+\frac{J_{\mu}^{2}}{2J_{l}^{2}}\left[\epsilon_{\mu,{\rm s}}^{(0)}-i\sqrt{4J_{l}^{2}-\left(\epsilon_{\mu,{\rm s}}^{(0)}\right)^{2}}\right], \label{sym}
\end{eqnarray}
and
\begin{eqnarray}
\epsilon_{\mu,{\rm as}}^{(2)}=\epsilon_{\mu,{\rm as}}^{(0)}+\frac{J_{\mu}^{2}}{2J_{l}^{2}}\left[\epsilon_{\mu,{\rm as}}^{(0)}-i\sqrt{4J_{l}^{2}-\left(\epsilon_{\mu,{\rm as}}^{(0)}\right)^{2}}\right]. \label{antisym}
\end{eqnarray}
The presence of a real-valued solution in the square roots imposes specific constraints to the allowed depth $V_{\mu}$. To observe significant particle jets under time-periodic driving, we need to be in the regime where the symmetric mode remains stable while the antisymmetric mode is damped, i.e., $\vert \epsilon_{\mu,{\rm s}}^{(0)}\vert>2J_{l}$ and $\vert \epsilon_{\mu,{\rm as}}^{(0)}\vert<2J_{l}$. As we specialize to negative trapping potentials $V_{\mu}=-|V_{\mu}|<0$, the inequalities yield $-J_{h}-2J_{l}<V_{\mu}<J_{h}-2J_{l}$.

\section{Nonlinear dynamics} \label{parametric}

In the following, we parametrically drive the system, and analyze the nonlinear dynamics by numerically solving Eqs.~(\ref{b0})-(\ref{linear}). We assume that the condensate is stable at $t<0$, with all the particles initially prepared in the central sites at the lowest symmetric mode before the periodic perturbation is turned on. Without any loss of generality, the antisymmetric mode is then seeded by taking $\beta\approx\gamma=1$ with a slight difference, and in the numerics we measure the energy in units of $J_{h}=1$, such that the times and the frequencies are measured in units of $1/J_{h}$ and $J_{h}$, respectively.

\subsection{Symmetric double-well potential}

\begin{figure}[htbp]
\includegraphics[width=\columnwidth]{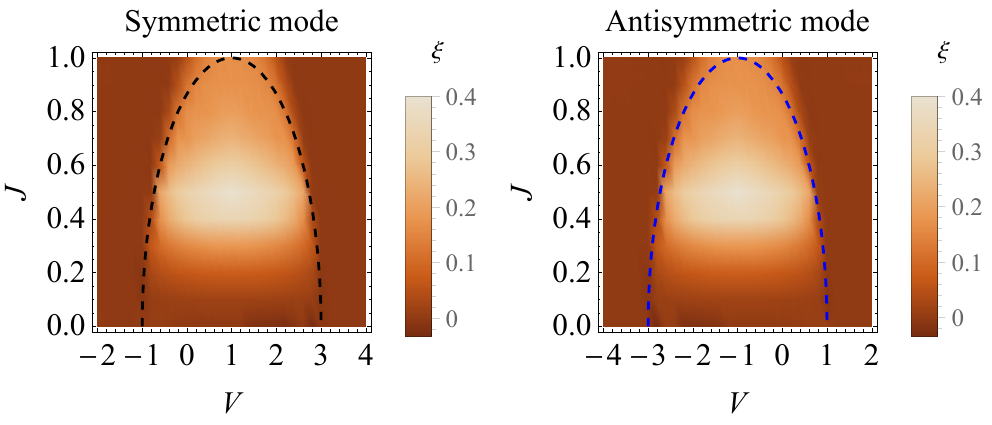}
\caption{Regimes for the symmetric mode and antisymmetric mode. The numerical results (color areas) are obtained by directly solving Eqs.~(\ref{b0}) and (\ref{c0}), while the analytical solutions (dashed lines) come from Eqs.~(\ref{sym}) and (\ref{antisym}). Color bars denotes the emission rate $\xi$ of the particles. Here, the tunneling strength is $J_{l}=1$, and the drive strength is $g=0.2$. We have taken the simulation time $\tau=50$.}
\label{regimeV}
\end{figure}

We first focus on the general case of a symmetric double well with equivalent depth ($V_{b}=V_{c}\equiv V$) and hopping amplitude ($J_{b}=J_{c}\equiv J$) to demonstrate the regimes for exciting the particles. The time evolution of atoms in the condensate is highly nonlinear, especially when the drive strength $g$ is large. One can, however, characterize the decay by fitting the total particle number in the central sites to an asymptotically exponential function as
\begin{eqnarray}
N_{0}(t)=\vert \varphi_{b,0}(t)\vert^{2}+\vert \varphi_{c,0}(t)\vert^{2} \approx \alpha e^{-\xi t},
\end{eqnarray}
where $\xi$ denotes the average emission rate of the particles from the condensate and $\alpha$ is a constant. As shown in Fig.~\ref{regimeV}, when the well depth is roughly $V<-1$, the symmetric mode remains quite stable for generic hopping amplitude $J$, while the antisymmetric mode can be unstable in the range of $-3<V<-1$, with the analytical solutions clearly delineating the boundaries. For subsequent analysis, we mainly take $V=-2$ with a relatively small $J$, and select appropriate driving parameters based upon the excitation regimes.

\begin{figure}[htbp]
\includegraphics[width=0.85\columnwidth]{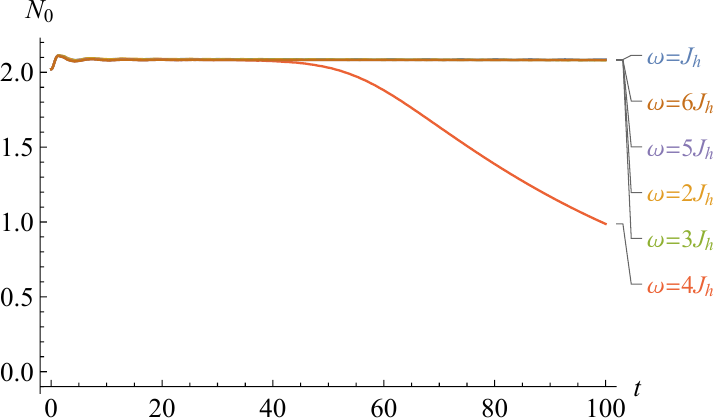}
\caption{Total particle number of the condensate $N_{0}$ as a function of time with different drive frequency $\omega$. Here, the depth of the well is $V=-2$, and the drive strength is $g=0.2$. The hopping amplitude is $J=0.1$, and the tunneling strength is $J_{l}=1$.}
\label{resofreq}
\end{figure}

Figure~\ref{resofreq} shows the short-time decay behavior of the total particle number $N_{0}$ under a fixed drive strength, with the drive frequency $\omega$ varying in multiples of $J_{h}$. It is clear that the system remains stable, unless the drive is resonant at $\omega=4J_{h}$. This arises from the fact that the wells support discrete bands corresponding to the local ground state (symmetric mode) and the local first excited state (antisymmetric mode), such that significant excitation occurs when the drive frequency matches the energy difference $\omega=2(\epsilon_{\mu,{\rm as}}^{(0)}-\epsilon_{\mu,{\rm s}}^{(0)})=4J_{h}$. Under this circumstance, pair atoms share half of the driving energy and are pumped to the excited state, after which they eject from the wells and propagate along the leads.

\begin{figure}[htbp]
\includegraphics[width=0.85\columnwidth]{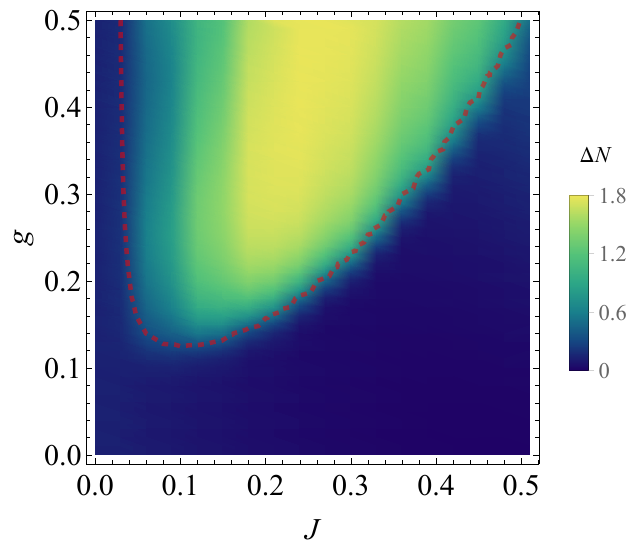}
\caption{Number of excited particles $\Delta N$ vs the drive strength $g$ and the hopping amplitude $J$. The color area results from the numerics of Eqs.~(\ref{b0}) and (\ref{c0}), and the red dashed line comes from the analytical solution in Eqs.~(\ref{ana1}) and (\ref{ana2}). We have taken the depth of the well $V=-2$ and the drive frequency $\omega=4J_{h}$. The tunneling strength is $J_{l}=1$, and the evolution time is $\tau=100$.}
\label{regimeJag}
\end{figure}

Since we focus on the regime of weak drive strength and weak hopping amplitude, the instabilities may be modified if the driving parameters are further increased. We thus define the number of excited particles
\begin{eqnarray}
\Delta N=N_{0}\left(t=0\right)-N_{0}\left(t=\tau\right),
\end{eqnarray}
and verify more quantitatively the interplay between the drive strength $g$ and the hopping amplitude $J$. As can be plainly seen in Fig.~\ref{regimeJag}, for generic hopping amplitude and weak drive strength, few particles can be ejected and $\Delta N$ remains small. In contrast, for stronger drives even a small hopping amplitude gives rise to significant emission. We see that a larger $J$ requires a stronger $g$, which can be described by the two-mode approximation
\begin{eqnarray}
b_{0}\left(t\right)&=&e^{-i (V-J_{h}) t} \chi(t) + e^{-i (V+J_{h}) t} \lambda(t), \\
c_{0}\left(t\right)&=&e^{-i (V-J_{h}) t} \chi(t) - e^{-i (V+J_{h}) t} \lambda(t),
\end{eqnarray}
where $\chi(t)$ and $\lambda(t)$ are the slowly varying amplitudes of symmetric and antisymmetric modes, respectively, and it yields (see Appendix~\ref{multiple scales})
\begin{eqnarray}
\partial_{t} \chi &=& -\frac{g}{2} \lambda^{2}\chi, \label{ana1} \\
\partial_{t} \lambda &=& -\frac{\Omega_{\rm as}}{2} \lambda + \frac{g}{2} \chi^{2}\lambda, \label{ana2}
\end{eqnarray}
with $\Omega_{\rm as}=-2J^{2}{\rm Im}\mathcal{G}_{11}\left(\epsilon_{\rm as}\right)$. This suggests that when $\Omega_{\rm{as}}>g\chi^{2}$, the particles remain bound and balanced in the two wells, while for $\Omega_{\rm{as}}<g\chi^{2}$ we get a buildup, followed by a significant particle emission. The threshold behavior appropriately delineates the numerics in Fig.~\ref{regimeJag}, and reflects the competition between dissipation induced by the leads and parametric amplification driven by interaction modulation. Notably, when the hopping amplitude is too small (roughly $J<0.05$), the excitation becomes greatly suppressed.

\subsection{Asymmetric double-well potential}

\begin{figure}[htbp]
\includegraphics[width=0.95\columnwidth]{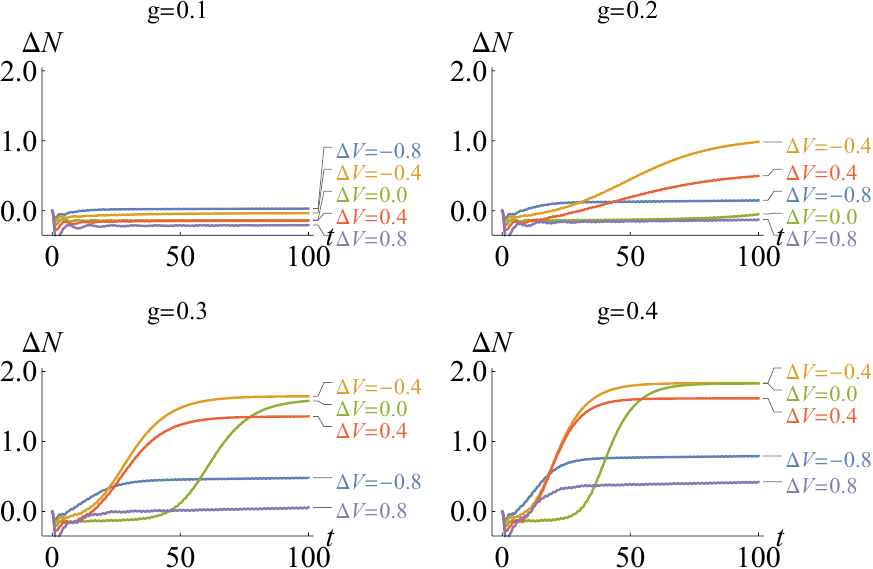}
\caption{Decay of the particle number $N_{0}$ under typical asymmetric double well $\Delta V$ for different drive strength $g$. Here, the depth of the left well is kept as $V_{b}=-2$, while the depth of the right one is varied as $V_{c}=-2.8$, $-2.4$, $-2$, $-1.6$ and $-1.2$, respectively. We have also taken the hopping amplitude $J_{b}=J_{c}=0.3$, the tunneling strength $J_{l}=1$, and the drive frequency $\omega=4$.}
\label{decayVg}
\end{figure}

We now turn to the asymmetric configuration with $V_{b}\neq V_{c}$. The depth asymmetry between the two wells, characterized by $\Delta V=V_{b}-V_{c}$, induces a relative on-site energy shift and consequently modifies the resonance conditions of the excitation spectrum. To ensure a consistent comparison, we choose parameters based upon the distinct instabilities identified in Fig.~\ref{regimeJag} and explore how the nonlinear dynamics develop. Figure~\ref{decayVg} shows the influence of the depth asymmetry on collective particle emission at a fixed hopping amplitude. When the drive strength is as small as $g=0.1$, which lies outside the parametric instability regions, the total particle number in the central sites exhibits only bounded oscillations for different values of $\Delta V$, and the number of ejected particles remains negligible, i.e., $|\Delta N|\approx0$.

For a stronger drive $g=0.2$, a relatively large depth asymmetry, either positive or negative (exemplified by $\Delta V=\pm0.8$), gives rise to only a weak enhancement of the emission, as the pronounced energy offset between the wells introduces a strong effective detuning, which reduces coherent tunneling and shifts the system away from optimal parametric resonance conditions. By contrast, for somewhat smaller asymmetries (e.g., $\Delta V=\pm0.4$), the decay can be significantly enhanced. In this regime, moderate asymmetry hybridizes the symmetric and antisymmetric modes, modifying their eigenfrequencies and effectively improving the resonance condition under parametric driving, and thereby the emission process becomes more efficient. When the drive strength is further increased to $g=0.3$ and $0.4$, the symmetric case ($\Delta V=0$) exhibits very weak decay at short times, followed by the emergence of a large pulse at intermediate driving intervals, and the excited number $\Delta N$ increases rapidly. For large values of depth asymmetry ($|\Delta V|=0.8$), the emission remains suppressed due to strong detuning, while the case with $|\Delta V|=0.4$ shows explicit enhancement, where the buildup stages becomes much shorter and the excited particles saturate faster than that of $\Delta V=0$.

\begin{figure}[htbp]
\includegraphics[width=0.95\columnwidth]{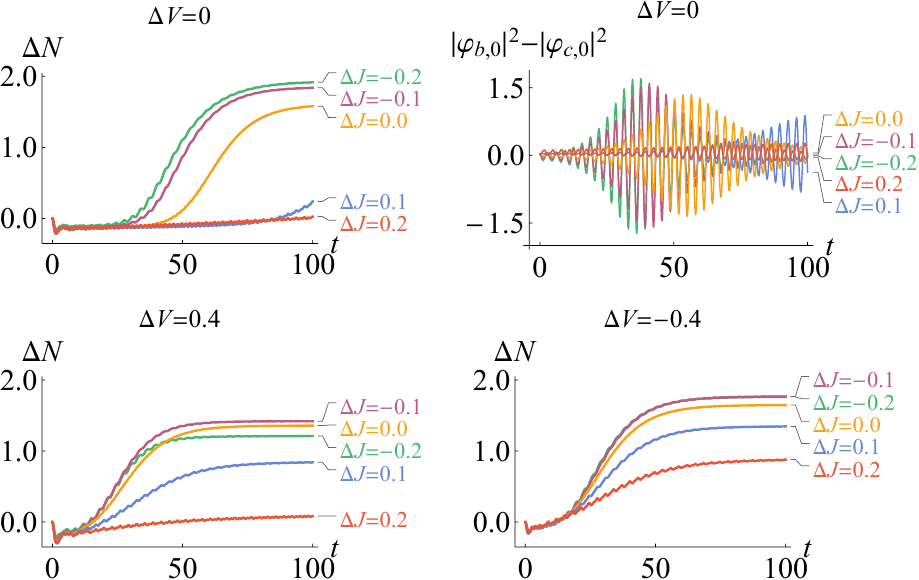}
\caption{Time evolution of the excited particle number under typical depth asymmetry $\Delta V$, with the hopping imbalance $\Delta J$ ranging from $-0.2$ to $0.2$. Accordingly, the depth of the wells are $V_{c}=-2$, and $V_{b}=-2$, $-1.6$, and $-2.4$, respectively. We have kept the hopping amplitude $J_{c}=0.3$, and the other parameters are $J_{l}=1$, $g=0.3$, and $\omega=4$.}
\label{dVtot}
\end{figure}

One can further explore how to control the excitation dynamics by tuning the hopping imbalance $\Delta J=J_{b}-J_{c}$ at a fixed drive strength. As shown in Fig.~\ref{dVtot}, for the case of $\Delta V=0$ negative hopping imbalance (e.g., $\Delta J=-0.1$ and $-0.2$) enhances the particle emission and shortens the buildup stage, while positive $\Delta J$ leads to insignificant excitations within the driving interval. Since $J_{c}$ is kept fixed, a negative $\Delta J$ implies a relative smaller amplitude $J_{b}$, which breaks the dynamical balance between the two wells and modifies the effective coupling of the unstable modes to the emission channel, and hence facilitates particle outflow once parametric instability sets in. To further elucidate the behavior, we calculate the particle imbalance $|\varphi_{b,0}|^{2}-|\varphi_{c,0}|^{2}$, where for $\Delta J=0.0$, $-0.1$ and $-0.2$ the imbalance first increases and subsequently decreases, indicating a buildup followed by particle emission and decay. In contrast, for $\Delta J=0.1$ and $0.2$ the imbalance keeps growing throughout, where the system maintains in the stage of buildup with only weak emission. As for the case of $|\Delta V|=0.4$, the interplay between energy detuning and hopping imbalance further enhances the instability. As a consequence, the emission rates can be increased under different $\Delta J$ with shorter buildup intervals when compared with that of $\Delta V=0$, and the excited particles gradually saturate.

\begin{figure}[htbp]
\includegraphics[width=0.9\columnwidth]{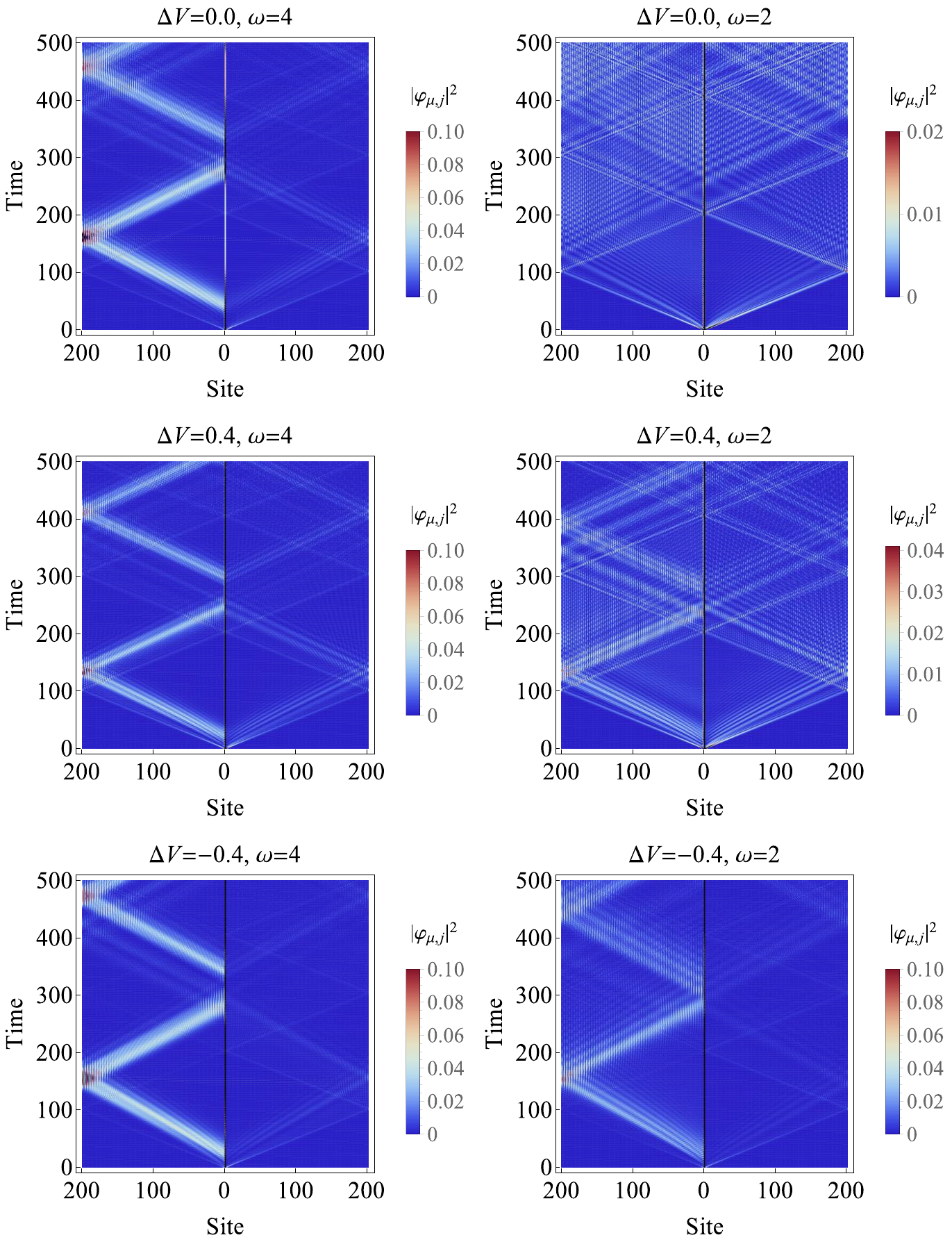}
\caption{Number of particles $|\varphi_{\mu,j}|^{2}$ on the $j$th site of each lead as a function of time, when the drive frequency is tuned to $\omega=4$ and $\omega=2$. The depth of the left well is kept as $V_{c}=-2$, while the right one varies as $V_{b}=-2$, $-1.6$, and $-2.4$, respectively. The other parameters are $g=0.3$, $J_{b}=0.1$, $J_{c}=0.3$, and $J_{l}=1$.}
\label{jetarray}
\end{figure}

Finally, we illustrate the structure of the jets by plotting the density distribution of particles on each lead. Since the periodic driving also provides energy in multiples, there can be another drive frequency $\omega=2$ resulting in significant particle emission. We compare the characteristics of particle jets under the two frequencies $\omega=4$ and $\omega=2$ for different degrees of depth asymmetry at a fixed hopping imbalance, as shown in Fig.~\ref{jetarray}. For the case of $\Delta V=0$ with $\omega=4$, most of the excited particles move towards the left after the buildup stage due to the hopping imbalance, and then propagate along the left lead. Upon reaching boundary sites, they reverse the direction of motion and gradually return to the central sites, repeating the emission behaviors over time. As for $\Delta V=0$ and $\omega=2$ the propagation appears similarly, yet with much fewer particles ejected. When the depth asymmetry is tuned to $\Delta V=0.4$, the emission with $\omega=4$ is slightly reduced, while $\omega=2$ leads to larger particle jets when compared with that of $\Delta V=0$. With respect to $\Delta V=-0.4$, both frequencies result in the enhancement of the visible pulses.

\section{Conclusions}\label{conclusion}

We have considered a one-dimensional infinite lattice in which a Bose-Einstein condensate is confined by a double-well potential, and external time-periodic driving is applied to modulate the interatomic interactions. We focus on the nonequilibrium dynamics of collective particle emission from the central sites into the leads, and analyze the roles of the depth asymmetry and the hopping imbalance.

For a symmetric double-well configuration, significant particle jets can be observed when the drive frequency resonates with the energy difference between the local ground state and the first excited state, and the interplay between the drive strength and the hopping amplitude explicitly determines the excitation regimes. When a depth asymmetry between the wells is introduced, moderate biases substantially enhance the emission by improving the resonance condition, whereas excessive asymmetry leads to its suppression due to detuning-induced localization. Further control over the particle jets can be achieved by tuning the hopping amplitudes between the wells and the leads. A finite hopping imbalance modifies the coupling of the unstable modes to the emission channels, thereby altering the emission rate. One can thus, in a specific experiment, manipulate both the intensity and directionality of the jets through the adjustments of either the depth asymmetry or the hopping imbalance, which contributes to the understanding of symmetry-breaking effects in driven quantum systems.

\section*{Acknowledgements}
This work was supported by National Natural Science Foundation of China (Grant No.~12505022) and the Natural Science Research Start-up Foundation of Recruiting Talents of Nanjing University of Posts and Telecommunications (Grant No.~NY223065). Z.L. received support from Nanhang Jincheng College (Grant Nos.~XJ2025003 and 2025JCXY62).


\begin{appendix}

\section{Finite interaction $U$}
\label{finiteU}
\renewcommand{\thefigure}{A\arabic{figure}}
\setcounter{figure}{0}
We consider the general case of $V_{b}=V_{c}\equiv V$ with the finite on-site interaction strength $U$ being included and the periodic driving being absent. Based upon Eqs.~(\ref{b0}) and (\ref{c0}) and the stationary ansatz with $|\beta|=|\gamma|\equiv\xi$, we obtain the similar energy levels with respect to the symmetric mode and antisymmetric mode, respectively,
\begin{eqnarray} \label{symU}
\epsilon_{\rm s}^{(2)} &=& \left(V+U\xi^{2}-J_{h}\right)+\frac{J_{\mu}^{2}}{2J_{l}^{2}}\left[\left(V+U\xi^{2}-J_{h}\right) \right. \nonumber \\
&&\left.-i\sqrt{4J_{l}^{2}-\left(V+U\xi^{2}-J_{h}\right)^{2}}\right],
\end{eqnarray}
and
\begin{eqnarray}\label{antisymU}
\epsilon_{\rm as}^{(2)} &=& \left(V+U\xi^{2}+J_{h}\right)+\frac{J_{\mu}^{2}}{2J_{l}^{2}}\left[\left(V+U\xi^{2}+J_{h}\right) \right. \nonumber \\
&&\left.-i\sqrt{4J_{l}^{2}-\left(V+U\xi^{2}+J_{h}\right)^{2}}\right].
\end{eqnarray}
The square roots $\mathcal{A}_{\rm s}=\sqrt{4J_{l}^{2}-\left(V+U\xi^{2}-J_{h}\right)^{2}}$ and $\mathcal{A}_{\rm as}=\sqrt{4J_{l}^{2}-\left(V+U\xi^{2}+J_{h}\right)^{2}}$ explicitly determine the excitation regimes, and basically the available $V$ varies with different $U$.

\begin{figure}[htbp]
\includegraphics[width=0.8\columnwidth]{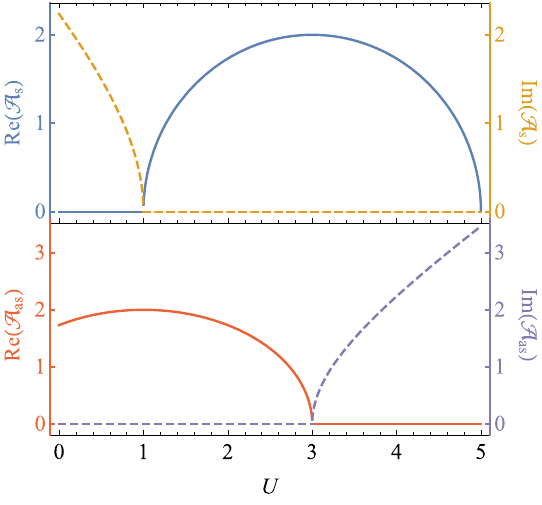}
\caption{$\mathcal{A_{\rm s}}$ and $\mathcal{A_{\rm as}}$ vs $U$ for $V=-2$, $J_{h}=1$, and $\xi=1$.}
\label{figA1}
\end{figure}

According to the typical parameters in the main text, one can verify the properties of $\mathcal{A}_{\rm s}$ and $\mathcal{A}_{\rm as}$ from Fig.~\ref{figA1}: For large on-site interaction strength $U>3$, the antisymmetric mode keeps stable and the particle emission can be significantly suppressed (i.e., the self-trapping), while at moderate interactions $1<U<3$ nonlinear excitations might become prominent and higher-order effects appear, which falls beyond the primary scope of the present work. We are particularly interested in the scenario where the symmetric mode remains stable and the antisymmetric mode is damped (roughly for $U<1$), and thus in the calculations we specifically work at the limit of $U=0$ to largely simplify the analysis.

\section{Frequency-domain Green's function}
\label{Green's function}
Here we present the derivation of the frequency-domain Green's function $\mathcal{G}_{11}$ in the main text. We begin from the matrix form of equation of motion for $j\geq1$ in Eq.~(\ref{linear}),
\begin{eqnarray}
i\frac{\partial}{\partial t}\left(
\begin{array}{c}
\varphi_{\mu,1}  \\
\varphi_{\mu,2}  \\
\varphi_{\mu,3}  \\
\varphi_{\mu,4}  \\
\vdots
\end{array}
\right)
&=&\left(
\begin{array}{ccccc}
0 & -J_{l} & 0 & 0 & \cdots  \\
-J_{l} & 0 & -J_{l} & 0 & \cdots  \\
0 & -J_{l} & 0 & -J_{l} & \cdots  \\
0 & 0 & -J_{l} & 0 & \cdots  \\
\vdots  & \vdots  & \vdots  & \vdots  & \ddots
\end{array}
\right) \left(
\begin{array}{c}
\varphi_{\mu,1}  \\
\varphi_{\mu,2}  \\
\varphi_{\mu,3}  \\
\varphi_{\mu,4}  \\
\vdots
\end{array}
\right) \nonumber \\
&&-J_{l}\left(
\begin{array}{c}
\varphi_{\mu,0}  \\
0 \\
0 \\
0 \\
\vdots
\end{array}
\right).
\end{eqnarray}
Using the Fourier transform $\varphi_{\mu,j}\left(\omega\right) =\int dt e^{-i\omega t}\varphi_{\mu,j}\left(t\right)$, in the frequency domain we get
\begin{eqnarray}
\left(
\begin{array}{c}
\varphi_{\mu,1}\left( \omega \right)  \\
\varphi_{\mu,2}\left( \omega \right)  \\
\varphi_{\mu,3}\left( \omega \right)  \\
\varphi_{\mu,4}\left( \omega \right)  \\
\vdots
\end{array}%
\right)  &=&\left(
\begin{array}{ccccc}
\omega & J_{l} & 0 & 0 & \cdots  \\
J_{l} & \omega & J_{l} & 0 & \cdots  \\
0 & J_{l} & \omega & J_{l} & \cdots  \\
0 & 0 & J_{l} & \omega & \cdots  \\
\vdots  & \vdots  & \vdots  & \vdots  & \ddots
\end{array}
\right) ^{-1}\left(
\begin{array}{c}
-J_{l}\varphi_{\mu,0}\left( \omega \right)  \\
0 \\
0 \\
0 \\
\vdots
\end{array}
\right)  \nonumber \\
&=&\left(
\begin{array}{ccccc}
\mathcal{G}_{11} & \mathcal{G}_{12} & \mathcal{G}_{13}  & \cdots  \\
\mathcal{G}_{21} & \mathcal{G}_{22} & \mathcal{G}_{23}  & \cdots  \\
\mathcal{G}_{31} & \mathcal{G}_{32} & \mathcal{G}_{33}  & \cdots  \\
\mathcal{G}_{41} & \mathcal{G}_{42} & \mathcal{G}_{43}  & \cdots  \\
\vdots  & \vdots  & \vdots   & \ddots
\end{array}
\right) \left(
\begin{array}{c}
-J_{l}\varphi_{\mu,0}\left( \omega \right)  \\
0 \\
0 \\
0 \\
\vdots
\end{array}
\right)  \nonumber \\
&=&\left(
\begin{array}{c}
-J_{l}\varphi_{\mu,0}\left( \omega \right) \mathcal{G}_{11} \\
-J_{l}\varphi_{\mu,0}\left( \omega \right) \mathcal{G}_{21} \\
-J_{l}\varphi_{\mu,0}\left( \omega \right) \mathcal{G}_{31} \\
-J_{l}\varphi_{\mu,0}\left( \omega \right) \mathcal{G}_{41} \\
\vdots
\end{array}
\right),
\end{eqnarray}
with $\mathcal{G}_{mn}$ being the element of the inverse matrix, and a general form for $\varphi_{\mu,j}$ is thus straightforward,
\begin{equation} \label{muj}
\varphi_{\mu,j}(\omega)=-J_{l} \varphi_{\mu,0}(\omega) \mathcal{G}_{j1}(\omega).
\end{equation}
We assume that $\varphi_{\mu,j}\left( \omega \right) =\eta e^{-\kappa \left( \omega \right) j}$. For
$j=1$,
\begin{equation}
\omega \eta e^{-\kappa (\omega)}=-J_{l}\left[\eta e^{-2 \kappa(\omega)}+\varphi_0(\omega)\right],
\end{equation}
from which we obtain
\begin{equation} \label{eta}
\eta=\frac{-J_{l} e^{\kappa(\omega)}}{\omega+J e^{-\kappa(\omega)}},
\end{equation}
while for $j>1$,
\begin{eqnarray}
\omega e^{-\kappa \left( \omega \right) j} &=&-J_{l}\left[ e^{-\kappa \left(
\omega \right) \left( j+1\right) }+ e^{-\kappa \left( \omega \right) \left(
j-1\right) }\right],
\end{eqnarray}
i.e.,
\begin{equation} \label{kappa}
2\cosh \kappa\left( \omega \right)  =-\frac{\omega}{J_{l}}.
\end{equation}
The value of $\kappa$ typically depends on the driving frequency. For $|\omega|>2J_{l}$, $\kappa$ is real, giving rise to evanescent modes in the leads and hence bound-state behavior without dissipation. As for $|\omega|<2J_{l}$, the complex $\kappa$ corresponds to propagating modes in the leads. In this regime, the leads act as a continuum bath, allowing particles to escape to infinity and thus leading to distinct particle jets. Inserting Eq.~(\ref{kappa}) into Eq.~(\ref{eta}) that we immediately have $\lambda=\varphi_0(\omega)$, and according to Eq.~(\ref{muj}) we can reach
\begin{equation}
\mathcal{G}_{j1}\left( \omega \right) =-\frac{e^{-\kappa \left( \omega \right) j}}{J_{l}}
\end{equation}
with
\begin{equation}
\kappa \left( \omega \right) ={\rm Arc}\left[\cosh \left( -\frac{\omega }{2J_{l}}\right) \right].
\end{equation}
As for $\mathcal{G}_{11}\left( \omega \right)$, we have the quadratic equation
\begin{eqnarray}
J_{l}^{2}\mathcal{G}_{11}(\omega)+\frac{1}{\mathcal{G}_{11}(\omega)} &=&-J_{l}e^{-\kappa \left( \omega \right)
}-J_{l}e^{\kappa \left( \omega \right) },
\end{eqnarray}
which yields
\begin{eqnarray}
\mathcal{G}_{11}(\omega)&=&\frac{\omega}{2J_{l}^{2}}-i \sqrt{\frac{1}{J_{l}^{2}}-\frac{\omega^{2}}{4J_{l}^{4}}}.
\end{eqnarray}

\section{Method of multiple scales}
\label{multiple scales}
Based on the two-mode approximation, we utilize the method of multiple scales and make the ansatz
\begin{eqnarray}
b_{0}\left(t\right)&=&e^{-i (V-J_{h}) t} \chi(t) + e^{-i (V+J_{h}) t} \lambda(t), \\
c_{0}\left(t\right)&=&e^{-i (V-J_{h}) t} \chi(t) - e^{-i (V+J_{h}) t} \lambda(t).
\end{eqnarray}
We also use the notation $\epsilon_{\rm {s}}=V-J_{h}$, $\epsilon_{\rm {as}}=V+J_{h}$ and $J_{\mu}\equiv J$ in the following, and the equations of motion for $\chi$ and $\lambda$ are thus straightforward
\begin{eqnarray}
ie^{i\epsilon_{\rm{s}}t}\partial_{t}\frac{b_{0}+c_{0}}{2} &=& \left(\epsilon_{\rm{s}}+i\partial_{t}\right)\chi \nonumber \\
&=& \left[V-J_{h}+J^{2}\mathcal{G}_{11}\left(\epsilon_{\rm{s}}\right)\right]\chi \nonumber \\
&&+\frac{e^{i\epsilon_{\rm{s}}t}}{2}g\left(t\right)\left(\vert b_{0}\vert^{2}b_{0}+\vert c_{0}\vert^{2}c_{0}\right), \\
ie^{i\epsilon_{\rm{as}}t}\partial_{t}\frac{b_{0}-c_{0}}{2} &=& \left(\epsilon_{\rm{as}}+i\partial_{t}\right)\lambda \nonumber \\
&=& \left[V+J_{h}+J^{2}\mathcal{G}_{11}\left(\epsilon_{\rm{as}}\right)\right]\lambda \nonumber \\
&&+\frac{e^{i\epsilon_{\rm{as}}t}}{2}g\left(t\right)\left(\vert b_{0}\vert^{2}b_{0}-\vert c_{0}\vert^{2}c_{0}\right).
\end{eqnarray}
There will be a resonance when the drive frequency is tuned to $\omega=2\left(\epsilon_{\rm{as}}-\epsilon_{\rm{s}}\right)$, and we keep only the resonant terms, such that
\begin{eqnarray}
\frac{e^{i\epsilon_{\rm{s}} t}}{2}g \sin\left(\omega t\right) \left(\vert b_{0}\vert^{2} b_{0}+\vert c_{0}\vert^{2} c_{0}\right) &\approx&
\frac{g}{2i} \lambda^{2} \chi^{*}, \\
\frac{e^{i\epsilon_{\rm{as}} t}}{2}g \sin\left(\omega t\right) \left(\vert b_{0}\vert^{2} b_{0}-\vert c_{0}\vert^{2} c_{0}\right) &\approx&
\frac{-g}{2i} \chi^{2} \lambda^{*},
\end{eqnarray}
and the equations of motion become
\begin{eqnarray}
i\partial_{t}\chi &=& J_h^{2} \mathcal{G}_{11}\left(\epsilon_{\rm{s}}\right) \chi +\frac{g}{2i} \lambda^{2} \chi^{*},  \\
i\partial_{t}\lambda &=& J_h^{2} \mathcal{G}_{11}\left(\epsilon_{\rm as}\right) \lambda -\frac{g}{2 i} \chi^{2} \lambda^{*},
\end{eqnarray}
which leads to
\begin{eqnarray}
\partial_{t} \vert\chi\vert^{2} &=& -\Omega_{\rm s}\vert\chi\vert^{2}-\frac{g}{2}  \left[\left(\chi^{*} \lambda\right)^{2} + \left(\lambda^{*}\chi\right)^{2}\right],   \\
\partial_{t} \vert\lambda\vert^{2} &=& -\Omega_{\rm as} \vert\lambda\vert^{2} +\frac{g}{2} \left[\left(\chi^{*} \lambda\right)^{2}+ \left(\lambda^{*}\chi\right)^{2}\right],
\end{eqnarray}
where $\Omega_{\rm s}=-2J^{2}{\rm Im}\mathcal{G}_{11}\left(\epsilon_{\rm s}\right)$ and $\Omega_{\rm as}=-2J^{2}{\rm Im}\mathcal{G}_{11}\left(\epsilon_{\rm as}\right)$. Since we seed the system in the symmetric mode $\Omega_{s}=0$, a simplification of treating $\chi$ and $\lambda$ to be real yields
\begin{eqnarray}
\partial_{t} \chi &=& -\frac{g}{2} \lambda^{2}\chi, \\
\partial_{t} \lambda &=& -\frac{\Omega_{\rm as}}{2} \lambda + \frac{g}{2} \chi^{2}\lambda.
\end{eqnarray}
These equations can readily be solved numerically, and the total particle number and particle imbalance are approximated, respectively,
\begin{eqnarray}
\vert b_{0}\vert^{2}+\vert c_{0}\vert^{2} &=& 2 \left(\chi^{2}+\lambda^{2}\right), \\
\vert b_{0}\vert^{2}-\vert c_{0}\vert^{2} &=& 4 \chi \lambda \cos\left(2J_{h}t\right).
\end{eqnarray}

\end{appendix}


\begin{thebibliography}{99}


\bibitem{bloch} I. Bloch, J. Dalibard, and W. Zwerger, Many-body physics with ultracold gases, \href{https://doi.org/10.1103/RevModPhys.80.885}{Rev. Mod. Phys. {\bf 80}, 885 (2008).}

\bibitem{chin} C. Chin, R. Grimm, P. Julienne, and E. Tiesinga, Feshbach resonances in ultracold gases, \href{https://doi.org/10.1103/RevModPhys.82.1225}{Rev. Mod. Phys. {\bf 82}, 1225 (2010).}


\bibitem{gati} R. Gati and M. K. Oberthaler, A bosonic Josephson junction, \href{https://doi.org/10.1088/0953-4075/40/10/R01}{J. Phys. B: At. Mol. Opt. Phys. {\bf 40}, R61 (2007).}

\bibitem{walls} G. J. Milburn, J. Corney, E. M. Wright, and D. F. Walls, Quantum dynamics of an atomic Bose-Einstein condensate in a double-well potential, \href{https://doi.org/10.1103/PhysRevA.55.4318}{Phys. Rev. A {\bf 55}, 4318 (1997).}

\bibitem{ananikian} D. Ananikian and T. Bergeman, Gross-Pitaevskii equation for Bose particles in a double-well potential: Two-mode models and beyond, \href{http://doi.org/10.1103/PhysRevA.73.013604}{Phys. Rev. A {\bf 73}, 013604 (2006).}

\bibitem{smerzi0} S. Giovanazzi, A. Smerzi, and S. Fantoni, Josephson Effects in Dilute Bose-Einstein Condensates, \href{https://doi.org/10.1103/PhysRevLett.84.4521}{Phys. Rev. Lett. {\bf 84}, 4521 (2000).}

\bibitem{shin1}Y. Shin, M. Saba, A. Schirotzek, T. A. Pasquini, A. E. Leanhardt, D. E. Pritchard, and W. Ketterle, Distillation of Bose-Einstein Condensates in a Double-Well Potential, \href{https://doi.org/10.1103/PhysRevLett.92.150401}{Phys. Rev. Lett. {\bf 92}, 150401 (2004).}

\bibitem{giovanazzi}  S. Giovanazzi, J. Esteve, and M. K. Oberthaler, Effective parameters for weakly coupled Bose-Einstein condensates, \href{http://doi.org/10.1088/1367-2630/10/4/045009}{New J. Phys. {\bf 10}, 045009 (2008).}

\bibitem{diaz} B. Juli\'{a}-D\'{i}az, J. Martorell, M. Mel\'{e}-Messeguer, and A. Polls, Beyond standard two-mode dynamics in bosonic Josephson junctions, \href{http://doi.org/10.1103/PhysRevA.82.063626}{Phys. Rev. A {\bf 82}, 063626 (2010).}

\bibitem{qian} Y. Y. Qian, M. Gong, and C. W. Zhang, Quantum transport of bosonic cold atoms in double-well optical lattices, \href{http://doi.org/10.1103/PhysRevA.84.013608}{Phys. Rev. A {\bf 84}, 013608 (2011).}

\bibitem{susanto} H. Susanto, J. Cuevas, and P. Krüger, Josephson tunnelling of dark solitons in a double-well potential, \href{http://doi.org/10.1088/0953-4075/44/9/095003}{J. Phys. B: At. Mol. Opt. Phys. {\bf 44}, 095003 (2011).}

\bibitem{zhang} D.-W. Zhang, L.-B. Fu, Z. D. Wang, and S.-L. Zhu, Josephson dynamics of a spin-orbit-coupled Bose-Einstein condensate in a double-well potential, \href{http://doi.org/10.1103/PhysRevA.85.043609}{Phys. Rev. A {\bf 85}, 043609 (2012).}

\bibitem{gu} H.-L. Zheng and Q. Gu, Dynamics of Bose-Einstein condensates in a one-dimensional optical lattice with double-well potential, \href{http://doi.org/10.1007/s11467-013-0321-0}{Front. Phys. {\bf 8}, 375 (2013).}

\bibitem{hou}  J. P. Hou, X.-W. Luo, K. Sun, T. Bersano, V. Gokhroo, S. Mossman, P. Engels, and C. W. Zhang, Momentum-Space Josephson Effects, \href{https://doi.org/10.1103/PhysRevLett.120.120401}{Phys. Rev. Lett. {\bf 120}, 120401 (2018).}

\bibitem{ying} Y.-J. Ying and H.-B. Li, Dynamics of Bose-Einstein condensation in an asymmetric double-well potential, \href{https://doi.org/10.7498/aps.72.20230419}{Acta Phys. Sin. {\bf 72}, 130303 (2023).}

\bibitem{sicks} J. Sicks and H. Rieger, Double-well Bose-Hubbard model with nearest-neighbor and cavity-mediated long-range interactions, \href{https://doi.org/10.1103/PhysRevA.109.033317}{Phys. Rev. A {\bf 109}, 033317 (2024).}

\bibitem{hamza} D. A. Hamza and J. Chwedeńczuk, Metrology using atoms in an array of double-well potentials, \href{https://doi.org/10.48550/arXiv.2507.11395}{arXiv:2507.11395.}

\bibitem{pitaevskii} L. Pitaevskii and S. Stringari, Thermal vs Quantum Decoherence in Double Well Trapped Bose-Einstein Condensates, \href{https://doi.org/10.1103/PhysRevLett.87.180402}{Phys. Rev. Lett. {\bf 87}, 180402 (2001).}

\bibitem{shin2} Y. Shin, M. Saba, T. A. Pasquini, W. Ketterle, D. E. Pritchard, and A. E. Leanhardt, Atom Interferometry with Bose-Einstein Condensates in a Double-Well Potential, \href{https://doi.org/10.1103/PhysRevLett.92.050405}{Phys. Rev. Lett. {\bf 92}, 050405 (2004).}

\bibitem{smerzi1} S. Raghavan, A. Smerzi, S. Fantoni, and S. R. Shenoy, Coherent oscillations between two weakly coupled Bose-Einstein condensates: Josephson effects, $\pi$ oscillations, and macroscopic quantum self-trapping, \href{https://doi.org/10.1103/PhysRevA.59.620}{Phys. Rev. A {\bf 59}, 620 (1999).}

\bibitem{salgueiro} A. N. Salgueiro, A.F.R. de Toledo Piza, G. B. Lemos, R. Drumond, M. C. Nemes, and M. Weidemüller, Quantum dynamics of bosons in a double-well potential: Josephson oscillations, self-trapping and ultralong tunneling times, \href{https://doi.org/10.1140/epjd/e2007-00224-4}{Eur. Phys. J. D {\bf 44}, 537 (2007).}

\bibitem{wang} W. Wang, L. B. Fu, and X. X. Yi, Effect of decoherence on the dynamics of Bose-Einstein condensates in a double-well potential, \href{https://doi.org/10.1103/PhysRevA.75.045601}{Phys. Rev. A {\bf 75}, 045601 (2007).}

\bibitem{adhikari} S. K. Adhikari, Josephson oscillation and induced collapse in an attractive Bose-Einstein condensate, \href{https://doi.org/10.1103/PhysRevA.72.013619}{Phys. Rev. A {\bf 72}, 013619 (2005).}

\bibitem{abad} M. Abad, M. Guilleumas, R. Mayol, F. Piazza, D. M. Jezek, and A. Smerzi, Phase slips and vortex dynamics in Josephson oscillations between Bose-Einstein condensates, \href{https://doi.org/10.1209/0295-5075/109/40005}{EPL {\bf 109}, 40005 (2015).}

\bibitem{nieuwkerk} Y. D. van Nieuwkerk, J. Schmiedmayer, and F. H. L. Essler, Josephson oscillations in split one-dimensional Bose gases, \href{https://doi.org/10.21468/SciPostPhys.10.4.090}{SciPost Phys. {\bf 10}, 090 (2021). }

\bibitem{gillet} J. Gillet, M. A. Garcia-March, Th. Busch, and F. Sols, Tunneling, self-trapping, and manipulation of higher modes of a Bose-Einstein condensate in a double well, \href{https://doi.org/10.1103/PhysRevA.89.023614}{Phys. Rev. A {\bf 89}, 023614 (2014).}

\bibitem{zibold} T. Zibold, E. Nicklas, C. Gross, and M. K. Oberthaler, Classical Bifurcation at the Transition from Rabi to Josephson Dynamics, \href{https://doi.org/10.1103/PhysRevLett.105.204101}{Phys. Rev. Lett. {\bf 105}, 204101 (2010).}

\bibitem{roy} R. Roy, B. Chakrabarti, and A. Trombettoni, Quantum dynamics of few dipolar bosons in a double-well potential, \href{https://doi.org/10.1140/epjd/s10053-022-00345-2}{Eur. Phys. J. D {\bf 76}, 24 (2022).}

\bibitem{smerzi2} A. Smerzi, S. Fantoni, S. Giovanazzi, and S. R. Shenoy, Quantum Coherent Atomic Tunneling between Two Trapped Bose-Einstein Condensates, \href{https://doi.org/10.1103/PhysRevLett.79.4950}{Phys. Rev. Lett. {\bf 79}, 4950 (1997).}

\bibitem{jezek} M. Abad, M. Guilleumas, R. Mayol, M. Pi, and D. M. Jezek, Phase slippage and self-trapping in a self-induced bosonic Josephson junction, \href{https://doi.org/10.1103/PhysRevA.84.035601}{Phys. Rev. A {\bf 84}, 035601 (2011).}

\bibitem{albiez} M. Albiez, R. Gati, J. F?lling, S. Hunsmann, M. Cristiani, and M. K. Oberthaler, Direct Observation of Tunneling and Nonlinear Self-Trapping in a Single Bosonic Josephson Junction, \href{https://doi.org/10.1103/PhysRevLett.95.010402}{Phys. Rev. Lett. {\bf 95}, 010402 (2005).}

\bibitem{zollner} S. Z?llner, H.-D. Meyer, and P. Schmelcher, Tunneling dynamics of a few bosons in a double well, \href{https://doi.org/10.1103/PhysRevA.78.013621}{Phys. Rev. A {\bf 78}, 013621 (2008).}

\bibitem{zollner1} S. Z?llner, H.-D. Meyer, and P. Schmelcher, Few-Boson Dynamics in Double Wells: From Single-Atom to Correlated Pair Tunneling, \href{https://doi.org/10.1103/PhysRevLett.100.040401}{Phys. Rev. Lett. {\bf 100}, 040401 (2008).}

\bibitem{zollner2} B. Chatterjee, I. Brouzos, S. Z?llner, and P. Schmelcher, Few-boson tunneling in a double well with spatially modulated interaction, \href{https://doi.org/10.1103/PhysRevA.82.043619}{Phys. Rev. A {\bf 82}, 043619 (2010).}

\bibitem{maraj} M. Maraj, J.-B. Wang, J.-S. Pan, and W. Yi, Interaction-modulated tunneling dynamics in a mixture of Bose-Einstein condensates, \href{https://doi.org/10.1140/epjd/e2017-80353-9}{Eur. Phys. J. D {\bf 71}, 300 (2017).}

\bibitem{lai0} L. Q. Lai, Y. B. Yu, and E. J. Mueller, Resonant enhancement of particle emission from a parametrically driven condensate in a one-dimensional lattice, \href{https://doi.org/10.1103/PhysRevA.106.033302}{Phys. Rev. A {\bf 106}, 033302 (2022).}

\bibitem{lai1} L. Q. Lai and Z. Li, Effects of drive imbalance on the particle emission from a Bose-Einstein condensate in a one-dimensional lattice, \href{https://doi.org/10.1088/1674-1056/ad1172}{Chin. Phys. B {\bf 33}, 030308 (2024).}

\bibitem{cataliotti} F. S. Cataliotti, S. Burger, C. Fort, P. Maddaloni, F. Minardi, A. Trombettoni, A. Smerzi, and M. Inguscio, Josephson Junction Arrays with Bose-Einstein Condensates, \href{https://doi.org/10.1126/science.1062612}{Science {\bf 293}, 843 (2001).}

\bibitem{morsch} O. Morsch and M. Oberthaler, Dynamics of Bose-Einstein condensates in optical lattices, \href{https://doi.org/10.1103/RevModPhys.78.179}{Rev. Mod. Phys. {\bf 78}, 179 (2006).}

\bibitem{mistakidis} S. I. Mistakidis, A. G. Volosniev, R. E. Barfknecht, T. Fogarty, Th. Busch, A. Foerster, P. Schmelcher, and N. T. Zinner, Few-body Bose gases in low dimensions—A laboratory for quantum dynamics, \href{https://doi.org/10.1016/j.physrep.2023.10.004}{Phys. Rep. {\bf 1042}, 1 (2023).}

\bibitem{schumm} T. Schumm, S. Hofferberth, L. M. Andersson, S. Wildermuth, S. Groth, I. Bar-Joseph, J. Schmiedmayer, and P. Krüger, Matter-wave interferometry in a double well on an atom chip, \href{https://doi.org/10.1038/nphys125}{Nat. Phys. {\bf 1}, 57 (2005).}

\bibitem{theocharis} G. Theocharis, P. G. Kevrekidis, D. J. Frantzeskakis, and P. Schmelcher, Symmetry breaking in symmetric and asymmetric double-well potentials, \href{https://doi.org/10.1103/PhysRevE.74.056608}{Phys. Rev. E {\bf 74}, 056608 (2006).}

\bibitem{hall} B. V. Hall, S. Whitlock, R. Anderson, P. Hannaford, and A. I. Sidorov, Condensate Splitting in an Asymmetric Double Well for Atom Chip Based Sensors, \href{https://doi.org/10.1103/PhysRevLett.98.030402}{Phys. Rev. Lett. {\bf 98}, 030402 (2007).}

\bibitem{jezek1} H. M. Cataldo and D. M. Jezek, Dynamics in asymmetric double-well condensates,  \href{http://dx.doi.org/10.1103/PhysRevA.90.043610}{Phys. Rev. A {\bf 90}, 043610 (2014).}


\bibitem{kim} S. J. Kim, H. Yu, S. T. Gang, D. Z. Anderson, and J. B. Kim, Controllable asymmetric double well and ring potential on an atom chip, \href{https://doi.org/10.1103/PhysRevA.93.033612}{Phys. Rev. A {\bf 93}, 033612 (2016).}

\bibitem{gavrilov} M. Gavrilov and J. Bechhoefer, Erasure without Work in an Asymmetric Double-Well Potential, \href{https://doi.org/10.1103/PhysRevLett.117.200601}{Phys. Rev. Lett. {\bf 117}, 200601 (2016).}

\bibitem{cosme} J. G. Cosme, M. F. Andersen, and J. Brand, Interaction blockade for bosons in an asymmetric double well, \href{https://doi.org/10.1103/PhysRevA.96.013616}{Phys. Rev. A {\bf 96}, 013616 (2017).}

\bibitem{haldar} S. K. Haldar and O. E. Alon, Many-body quantum dynamics of an asymmetric bosonic Josephson junction, \href{https://doi.org/10.1088/1367-2630/ab4315}{New J. Phys. {\bf 21}, 103037 (2019).}

\bibitem{lindberg} D. R. Lindberg, N. Gaaloul, L. Kaplan, J. R. Williams, D. Schlippert, P. Boegel, E.-M. Rasel, and D. I. Bondar, Asymmetric tunneling of Bose-Einstein condensates,  \href{https://doi.org/10.1088/1361-6455/acae50}{J. Phys. B: At. Mol. Opt. Phys. {\bf 56}, 025302 (2023).}

\bibitem{korshynska} K. Korshynska and S. Ulbricht, Generalized Josephson effect in an asymmetric double-well potential at finite temperatures, \href{https://doi.org/10.1103/PhysRevA.109.043321}{Phys. Rev. A {\bf 109}, 043321 (2024).}

\bibitem{trenkwalder} A. Trenkwalder, G. Spagnolli, G. Semeghini, S. Coop, M. Landini, P. Castilho, L. Pezzè, G. Modugno, M. Inguscio, A. Smerzi, and M. Fattori, Quantum phase transitions with parity-symmetry breaking and hysteresis, \href{https://doi.org/10.1038/nphys3743}{Nat. Phys. {\bf 12}, 826 (2016).}

\bibitem{rubbo} C. P. Rubbo, S. R. Manmana, B. M. Peden, M. J. Holland, and A. M. Rey, Resonantly enhanced tunneling and transport of ultracold atoms on tilted optical lattices, \href{https://doi.org/10.1103/PhysRevA.84.033638}{Phys. Rev. A {\bf 84}, 033638 (2011).}

\bibitem{sias} C. Sias, A. Zenesini, H. Lignier, S. Wimberger, D. Ciampini, O. Morsch, and E. Arimondo, Resonantly Enhanced Tunneling of Bose-Einstein Condensates in Periodic Potentials, \href{https://doi.org/10.1103/PhysRevLett.98.120403}{Phys. Rev. Lett. {\bf 98}, 120403 (2007).}

\bibitem{sias1} A. Zenesini, C. Sias, H. Lignier, Y. Singh, D. Ciampini, O. Morsch, R. Mannella, E. Arimondo, A. Tomadin, and S. Wimberger, Resonant tunneling of Bose-Einstein condensates in optical lattices, \href{https://doi.org/10.1088/1367-2630/10/5/053038}{New J. Phys. {\bf 10}, 053038 (2008).}

\bibitem{buyskikh} A. S. Buyskikh, L. Tagliacozzo, D. Schuricht, C. A. Hooley, D. Pekker, and A. J. Dale, Resonant two-site tunneling dynamics of bosons in a tilted optical superlattice, \href{https://doi.org/10.1103/PhysRevA.100.023627}{Phys. Rev. A {\bf 100}, 023627 (2019).}

\bibitem{alon} A. Bhowmik and O. E. Alon, Longitudinal and transversal resonant tunneling of interacting bosons in a two-dimensional Josephson junction, \href{https://doi.org/10.1038/s41598-021-04312-6}{Sci. Rep. {\bf 12}, 627 (2022).}

\bibitem{lai2} L. Q. Lai and Z. Li, Interference-induced suppression of particle emission from a Bose-Einstein condensate in lattice with time-periodic modulations, \href{https://doi.org/10.1088/1674-1056/ad607b}{Chin. Phys. B {\bf 33}, 100303 (2024).}

\bibitem{lai3} L. Q. Lai, Z. Li, Q. H. Liu, and Y. B. Yu,  Intermittent Emission of Particles from a Bose-Einstein Condensate in a 1D Lattice, \href{https://doi.org/10.1002/andp.202300365}{Ann. Phys. (Berlin) {\bf 536}, 2300365 (2024).}

\bibitem{clark} L. W. Clark, A. Gaj, L. Feng, and C. Chin, Collective emission of matter-wave jets from driven Bose–Einstein condensates, \href{https://doi.org/10.1038/nature24272}{Nature {\bf 551}, 356 (2017).}

\bibitem{clark1} H. Fu, L. Feng, B. M. Anderson, L. W. Clark, J. Z. Hu, J. W. Andrade, C. Chin, and K. Levin, Density Waves and Jet Emission Asymmetry in Bose Fireworks, \href{https://doi.org/10.1103/PhysRevLett.121.243001}{Phys. Rev. Lett. {\bf 121}, 243001 (2018).}





\end{thebibliography}
\end{document}